\documentclass[11pt]{article}
\pdfoutput=1

\usepackage{jheppub} 
\usepackage{slashed}
\usepackage[T1]{fontenc}
\usepackage{amsmath}
\usepackage{amssymb}
\usepackage{esint}
\usepackage{lmodern}
\usepackage{slashed}
\usepackage{dsdshorthand}
\usepackage{tikz} 
\usepackage[bottom]{footmisc}
\usepackage{mathdots}
\usetikzlibrary{arrows.meta} 
\usepackage{tabularx}
\usetikzlibrary{arrows}
\usepackage{subcaption}
\usepackage{float}

\def\bz{\bar{z}}

\newtheorem*{theorem*}{Theorem}



\title{Critical slowing down of black hole phase transition and universal dynamic scaling in AdS black holes}

\author[a]{Mozib Bin Awal}
\author[a,b]{and Prabwal Phukon}

\affiliation[a]{Department of Physics, Dibrugarh University - Assam, 786004, India}
\affiliation[b]{Theoretical Physics Division, Centre for Atmospheric Studies, Dibrugarh University, Assam}

\emailAdd{rs\_mozibbinawal@dibru.ac.in}
\emailAdd{prabwal@dibru.ac.in}

\abstract{We investigate the dynamical critical behaviour of black hole phase transitions in anti de Sitter spacetime by extending the stochastic framework of free energy landscape dynamics to Kerr AdS black holes. By analyzing the Langevin evolution of the entropy (in contrast to the horizon radius in the RN-AdS case) near criticality, we demonstrate that the system exhibits pronounced critical slowing down, characterized by a significant increase of the autocorrelation time as the critical point is approached. This behaviour is further confirmed by the lowest eigenvalue of the Fokker-Planck equation. By analysing the dynamics along different thermodynamic paths, including variations in temperature, pressure, and angular momentum, and considering both directions - towards and away from criticality, we find that the relaxation time obeys a robust scaling relation, $\tau=|\epsilon|^{-2/3}$ near criticality. The same scaling exponent is obtained for RN-AdS, Kerr-AdS, and Bardeen black holes, suggesting the existence of an underlying universal dynamical behaviour across distinct black hole systems. Our results establish a connection between the geometry of the free energy landscape, stochastic nonequilibrium dynamics, and universal critical phenomena in black hole thermodynamics.}

\begin{document} 

\maketitle
\section{Introduction}\label{sec1}
Black hole thermodynamics has emerged as a central framework in modern gravitational physics. The pioneering contributions of Bekenstein, Hawking, Bardeen and others~\cite{Phys,bekens,Hawking,Hawking2,Bardeen} have led to the understanding that black holes can be described within a consistent thermodynamic framework, with well-defined notions of temperature and entropy. These developments naturally raise a deeper question: beyond this formal analogy, do black holes exhibit genuine thermodynamic behaviour, including phase structure and critical phenomena analogous to those observed in conventional systems? In this context, phase transitions in black holes emerge as a particularly compelling aspect of their thermodynamic description. The earliest studies along these lines were carried out by P.~C.~W.~Davies~\cite{Davies} and P.~Hut~\cite{Hut}, who examined signatures of such transitions in gravitational settings. The thermodynamic description of Anti-de Sitter (AdS) black holes acquired renewed significance following Maldacena’s proposal of the AdS/CFT correspondence in 1997~\cite{Maldacena}. A notable conceptual advance was the reinterpretation of the cosmological constant $\Lambda$ as a pressure term, leading to an extended formulation of black hole thermodynamics. Within this framework, black holes exhibit a remarkably rich phase structure. In particular, their phase behaviour shows a close qualitative similarity to that of ordinary thermodynamic systems such as fluids~\cite{Kubiz,Hawkpage,Cai,Kastor,Dolan,Dolan2,Dolan3,Kubizna,Xu,Xu2,Zhang}. Although, conventionally, phase transitions in black hole were studied using the standard free energy, recently many interesting alternative approaches to black hole thermodynamics has emerged. Thermodynamic geometry \cite{Ruppeiner:2012uc,Miao:2017cyt,Guo:2019oad}, thermodynamic topology \cite{Wu:2022whe,Liu:2022aqt,Fan:2022bsq}, using Lyapunov exponents \cite{first,le,Awal} are just to name a few.

The concept of a free energy landscape plays a central role in the study of phase transitions across diverse systems in physics, chemistry, and biology, offering valuable insight into their underlying dynamics~\cite{protein,bio,Golden,noneq}. Several studies have proposed that the underlying kinetics of phase transitions can be probed by analysing their dynamics, in the presence of thermal fluctuations, through the corresponding probabilistic Fokker-Planck equation defined on the free energy landscape~\cite{p1,p2,p3,p4}. However, the application of this framework to black hole systems remained largely unexplored until recently. In particular, Li and Wang~\cite{Li} investigated the dynamical aspects of black hole phase transitions within the free energy landscape framework. Using this approach, they analysed the Hawking-Page transition in both Einstein and massive gravity. The formalism was subsequently extended to the study of small-large phase transitions in Reissner-Nordström AdS (RN-AdS) black holes \cite{Li2}, which exhibit a close analogy with the van der Waals liquid-gas system. Analysis based on numerical solutions of the Fokker-Planck equation reveals that stochastic transitions between small and large black hole states can occur in both forward and reverse directions. Subsequently, several studies have extended the formalism to various other black hole systems \cite{fp1,fp2,fp3,fp4,fp5,fp6}. In these approaches, the order parameter is treated as a stochastic variable evolving under thermal fluctuations, with its dynamics governed by a Langevin equation incorporating dissipative, driving, and noise terms, or equivalently by the associated Fokker--Planck equation describing the probabilistic evolution on the free energy landscape. These studies introduced a novel framework for probing the kinetics of black hole phase transitions through the structure of the free energy landscape. The roles of temperature and the free energy barrier height in governing these transitions were also analysed. 

More recently, the dynamical aspects of black hole phase transitions have been explored in greater detail. In particular, in Ref. \cite{slowing} it was shown that near the critical and spinodal points, the system exhibits a pronounced critical slowing down, characterized by a significant increase in the autocorrelation time and fluctuations of the order parameter. The authors modelled the phase transition dynamics by treating the order parameter as a stochastic variable evolving on the free energy landscape, with its evolution governed by a Langevin equation or, equivalently, by the associated Fokker-Planck equation \cite{slowing}. This behaviour was attributed to the flattening of the free energy landscape and further supported by the analysis of the lowest eigenvalue of the corresponding Fokker-Planck equation. In addition, a kinetic crossover in the supercritical regime, associated with a Widom line, was identified through dynamical observables.

Motivated by this, we extend the analysis of critical slowing down to rotating Kerr-AdS black holes. Unlike in Ref. \cite{slowing}, where the horizon radius is taken as the order parameter, we consider the entropy as the relevant dynamical variable and investigate its stochastic evolution near criticality. By analysing the Langevin dynamics and the associated Fokker-Planck equation, we demonstrate the emergence of critical slowing down in rotating black holes. We further examine the scaling behaviour of the relaxation time along different thermodynamic paths, including variations in temperature, pressure, and angular momentum, and compare the results across RN-AdS, Kerr-AdS, and Bardeen black holes. Our findings indicate a robust power-law behaviour near the critical point, suggesting a degree of universality in the underlying dynamics.

The paper is organised as follows. In section~\ref{sec2}, we outline the thermodynamic framework and construct the free energy landscape relevant for the analysis. In section~\ref{sec3}, we investigate the emergence of critical slowing down in Kerr-AdS black holes within the stochastic dynamics approach. Section~\ref{sec4} is devoted to the analysis of the scaling behaviour of the relaxation time and its possible universality across different black hole systems. Finally, in section~\ref{sec5} we summarise our results and discuss their implications.

\section{Thermodynamics and Phase Transition Kinetics for Kerr Black Holes}\label{sec2}
We start with the metric of the Kerr AdS black hole in Boyer-Lindquist coordinates
\begin{equation}\label{eq2.1}
		ds^2=-\frac{\Delta}{\rho^2}\left [ dt-\frac{a sin^2\theta d\varphi}{\Xi} \right ]^2+\frac{\rho^2}{\Delta} dr^2 +\frac{\rho^2}{\Sigma} d\theta^2+\frac{\Xi sin^2 \theta} {\rho^2}\left[a dt-\frac{r^2+a^2}{\Xi} d\varphi\right]^2
	\end{equation}
with $\Delta=(r^2+a^2)\left(1+\frac{r^2}{l^2}\right)-2m r$, $\Sigma=1-\frac{a^2}{l^2} cos^2\theta,  \rho^2=r^2++a^2cos^2\theta, \Xi=1-\frac{a^2}{l^2}$
where $m$ represents the mass parameter, $a$ is the angular momentum per unit mass and $l$ is the AdS length. The ADM mass $M$ and the angular momentum $J$ can be expressed as 
\begin{equation}\label{eq2.2}
		M=\frac{m}{\Xi^2} \hspace{0.5cm},\hspace{0.5cm} 	J=\frac{a m}{\Xi^2} 
	\end{equation}
By setting $\Delta_r(r=r_+)=0$, we get the mass parameter $m$ as
\begin{equation}\label{eq2.3}
	m=\frac{\left(a^2+r_+^2\right) \left(\frac{r_+^2}{l^2}+1\right)}{2 r_+}
	\end{equation}
The entropy for Kerr AdS black hole is expressed as
\begin{equation}\label{eq2.4}
	S=\frac{\pi (r_{+}^2+a^2)}{\Xi}
	\end{equation}
Now, using equations \ref{eq2.2} and \ref{eq2.4}, we can express the mass of the Kerr AdS black hole in terms of entropy as \cite{bd}
\begin{equation}\label{eq2.5}
		M=\frac{\sqrt{4 \pi ^3 J^2 S+4 \pi ^4 J^2+S^4+2 \pi  S^3+\pi ^2 S^2}}{2 \pi ^{3/2} \sqrt{S}}
	\end{equation}
where the AdS length $l$ is considered to be $1$.

\begin{figure}[ht!]
\centering
\includegraphics[width=0.5\textwidth]{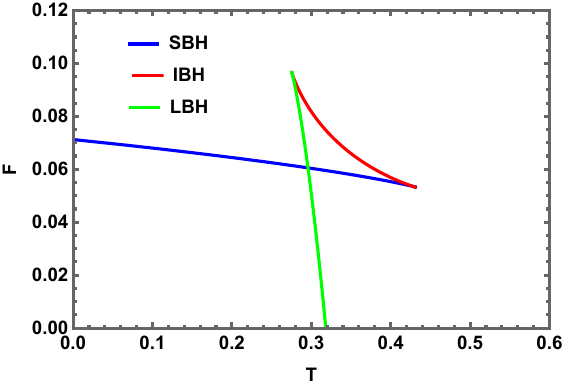}
\caption{Gibbs free energy (F) as a function of Hawking temperature. The swallow-tail behaviour indicates a first-order phase transition.}
\label{f1}
\end{figure}

The Kerr-AdS black hole in the canonical ensemble exhibits a phase structure analogous to that of the RN-AdS black hole, including a van der Waals-like first-order phase transition characterized by the swallow-tail behaviour of the Gibbs free energy, as shown in Fig.~\ref{f1}. The different branches of the free energy correspond to distinct black hole phases: the blue curve represents the small black hole (SBH) phase, the red curve corresponds to the intermediate black hole (IBH) phase, and the green curve denotes the large black hole (LBH) phase.

We now write down the generalised free energy for Kerr AdS black hole in canonical ensemble in terms of the black hole entropy as
\begin{equation}\label{eq2.6}
G(S)=\frac{\sqrt{4 \pi ^3 J^2 S+4 \pi ^4 J^2+S^4+2 \pi  S^3+\pi ^2 S^2}}{2 \pi ^{3/2} \sqrt{S}}-T S
\end{equation}
Similar to the RN-AdS case, Kerr-AdS black holes also undergo critical behaviour. The critical values of the relevant thermodynamic quantities can be determined from the conditions
\begin{equation}\label{eq2.7}
G^{(1)}=G^{(2)}=G^{(3)}=0
\end{equation}
where $G^{(n)}$ represents the $n$th order derivative with respect to the entropy $S$. Solving these equations, we obtain the following critical values for Kerr AdS black holes
\begin{equation}\label{eq2.8}
S_c=0.68732,\quad T_c=0.26986,\quad J_c=0.02393
\end{equation}
The spinodal points can be identified from the conditions that the first and second derivatives of the generalized free energy with respect to the entropy simultaneously vanish. Therefore, the phase transitions and thermodynamics can be completely analysed using the generalised free energy landscape.

The free energy landscape framework can also be utilized for investigating the kinetics of black hole phase transitions. The corresponding kinetic description is closely related to the time-dependent Ginzburg-Landau approach \cite{Hohenberg}, where the near-critical dynamics is governed by the dissipative evolution of a nonconserved order parameter~\cite{model1,model2}. 

Assuming that transitions between black hole states occur on the underlying free energy landscape, the stochastic evolution of the order parameter, identified in the present work with the black hole entropy, can be described by the Langevin equation
\begin{equation}\label{eq2.9}
\frac{d^2 S}{dt^2}
=
-\zeta \frac{dS}{dt}-\frac{\partial G}{\partial S}+\eta(t),
\end{equation}
Here, $\zeta$ denotes the dissipation coefficient, characterizing the interaction between the black hole degrees of freedom and the surrounding thermal environment. The stochastic term $\eta(t)$ is taken to be Gaussian white noise with vanishing mean, satisfying the fluctuation--dissipation relation
\begin{equation}\label{eq2.10}
\langle \eta(t) \rangle = 0,
\qquad
\langle \eta(t)\eta(s) \rangle = 2\zeta T\,\delta(t-s).
\end{equation}

Within this framework, the free energy landscape governs the deterministic component of the dynamics, while thermal fluctuations provide the stochastic contribution, together controlling the evolution of the system.

\section{Critical Slowing Down in Kerr AdS Black Holes}\label{sec3}
\subsection{Analytical Treatment}
A key feature of black hole phase transitions in the free energy landscape framework is the progressive flattening of the landscape as the system approaches the critical point. In this regime, the curvature of the free energy near its extrema decreases, reducing the effective restoring force that drives the system back to equilibrium. As a consequence, one may anticipate a significant slowing down of the dynamical evolution, since fluctuations can persist over longer timescales in a nearly flat landscape.

In general, such behaviour is identified as critical slowing down, which is typically associated with a divergence of the autocorrelation time or correlation length, along with enhanced fluctuations~\cite{Hohenberg}. The autocorrelation time $\tau$ is expected to exhibit a power-law scaling of the form $\tau \sim |\epsilon|^{-\Delta}$, where $\epsilon = (T - T_c)/T_c$ denotes the reduced temperature and $\Delta$ is the dynamic critical exponent. $\epsilon$ can also be defined in terms of the reduced pressure as $\epsilon = (P - P_c)/P_c$ or in terms of the angular momentum $J$ in a similar manner. This prolonged relaxation near criticality often serves as a precursor to phase transitions and is closely related to phenomena such as early-warning signals of critical transitions and the Kibble-Zurek mechanism~\cite{Kibble,Zurek}. In Ref. \cite{slowing}, the authors considered the horizon radius $r_+$ as the order parameter and then modelled the phase transition dynamics by treating $r_+$ as a stochastic variable evolving in the free energy landscape. In our study, we consider the entropy $S$ of the Kerr AdS black hole to be the order parameter. Although alternative choices of order parameter, such as the horizon radius, are possible, the resulting dynamical behaviour remains qualitatively unchanged as long as the chosen variable faithfully represents the underlying thermodynamic state and satisfies the same effective dissipative dynamics. This indicates that the observed features are not tied to a specific parametrization but reflect intrinsic properties of the system. In what follows in this section, we recapitulate what was done for the RN AdS case (with horizon radius as the order parameter) and generalize the analysis to Kerr AdS system

To make this connection explicit in the present context, we consider the deterministic evolution of the order parameter in the overdamped limit, governed by
\begin{equation}\label{eq3.1}
\zeta \frac{dS}{dt} = -\frac{dG}{dS}.
\end{equation}
Far from the criticality, the free energy landscape often takes the shape of a double-well or a single-well. In both the cases, one can approximate the free energy near a stationary point in the following way
\begin{equation}\label{eq3.2}
G(S) = G(S_e) + \frac{1}{2} G^{(2)}(S_e)\,(S - S_e)^2 + \cdots,
\end{equation}
where $S_e$ denotes the equilibrium value of the order parameter. At this point, the first derivative of the free energy vanishes, ensuring that the linear term in the expansion is absent.

For small perturbations around equilibrium, and neglecting higher-order contributions, the time evolution of the order parameter in the overdamped regime follows an exponential relaxation of the form
\begin{equation}\label{eq3.3}
S(t) - S_e = \bigl(S(0) - S_e\bigr)\,e^{-\gamma t},
\end{equation}
where $\gamma=\frac{G^2(S_e)}{\zeta}$ is called the decay factor and $S(0)$ is the initial value of the order parameter. This solution implies that the relaxation dynamics is associated with a characteristic timescale $\tau = 1/\gamma(T)$, determined by the ensemble temperature $T$. At the critical values, the free energy derivatives with respect to the black hole entropy are zero up to third order. The order parameter near the critical point evolves as 
\begin{equation}\label{eq3.4}
S(t) - S_c = \frac{S(0) - S_c}{\sqrt{(S(0) - S_c)^2\,\beta t + 1}},
\end{equation}
where $\beta = \frac{G^{(4)}(S_c)}{3\zeta}$, and, $\beta$ characterizes the strength of higher-order fluctuations of the order parameter in the vicinity of the critical point. It effectively captures the leading non-quadratic contribution to the free energy, reflecting the deviation of the landscape from a simple parabolic form.

We now turn to the stochastic evolution of the order parameter and analyse the kinetics near the critical point through the autocorrelation function. In this regime, we focus on fluctuations around a given stable state, assuming that the dynamics remains confined within its vicinity. We consider the stochastic dynamics of the order parameter in the overdamped limit which can be approximated by the Langevin equation
\begin{equation}\label{eq3.5}
\frac{dS}{dt} = -\frac{G^{(2)}(S_e)}{\zeta}\,(S - S_e) + \frac{1}{\zeta}\,\eta(t),
\end{equation}
where $\eta(t)$ represents the stochastic noise term.
In such a stochastic dynamical framework, signatures of critical slowing down can be identified through the divergence of the autocorrelation time of the order parameter as the system approaches criticality.

The standard solution to equation \ref{eq3.5} can be expressed as
\begin{equation}\label{eq3.6}
S(t) - S_e = \bigl(S(0) - S_e\bigr)e^{-t/\tau} + \frac{1}{\zeta} \int_0^t e^{-(t - t')/\tau}\,\eta(t')\,dt',
\end{equation}
which describes the combined effects of deterministic relaxation and stochastic fluctuations. By multiplying both sides by $(S(0) - S_e)$ and taking the ensemble average, one gets
\begin{equation}\label{eq3.7}
\left\langle (S(t)-S_e)(S(0)-S_e) \right\rangle
=
\left\langle (S(0)-S_e)^2 \right\rangle e^{-t/\tau}
+
\frac{1}{\zeta}
\int_0^t
e^{-(t-t')/\tau}
\left\langle \eta(t')(S(0)-S_e)\right\rangle dt'.
\end{equation}
Considering the mean of the stochastic noise to be zero, leaves no contribution from the second term, and we obtain, for the autocorrelation function of the order parameter
\begin{equation}\label{eq3.8}
\left\langle (S(t)-S_e)(S(0)-S_e) \right\rangle
\sim e^{-t/\tau},
\end{equation}
This exponential behaviour is consistent with the deterministic time evolution given in Eq.\ref{eq3.3}. The associated autocorrelation time,
$\tau = \frac{\zeta}{G^{(2)}(S_e)}$, is governed by the curvature of the free energy landscape near equilibrium. As the system approaches the critical point, the second derivative of the free energy tends to vanish, causing the autocorrelation time to grow significantly and eventually diverge, signalling the onset of critical slowing down. 

\begin{figure}[h!]
    \centering
    \begin{subfigure}[b]{0.45\textwidth}
        \centering
        \includegraphics[width=\textwidth]{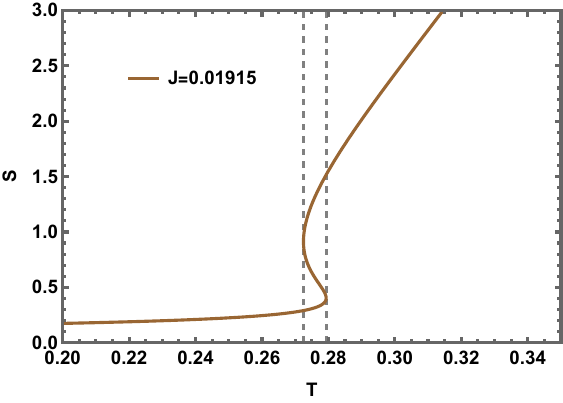}
        \caption{}
        \label{f2a}
    \end{subfigure}
    \hspace{0.8cm}
    \begin{subfigure}[b]{0.46\textwidth}
        \centering
        \includegraphics[width=\textwidth]{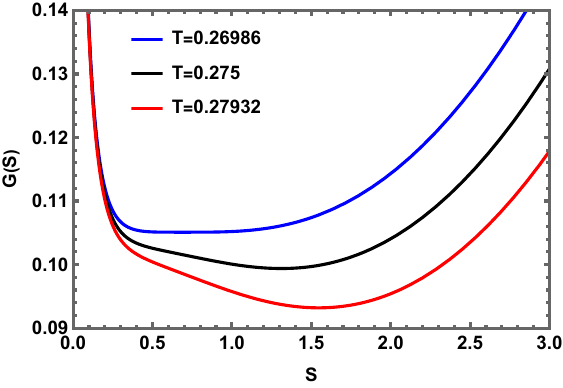}
        \caption{}
        \label{f2b}
    \end{subfigure}

    \caption{The left panel shows the temperature profile of the Kerr AdS black hole for $J<J_c$. Here the dashed line are the two spinodal temperatures. The right panel shows the free energy landscapes for different values of the Hawking temperature.}
    \label{f2}
\end{figure}

A similar behaviour also emerges near the spinodal point. In this regime, the generalized free energy develops features analogous to those near criticality. The quadratic approximation to the free energy ceases to remain valid since the derivatives of the free energy with respect to the order parameter vanish up to second order. As a result, the autocorrelation time $\tau$ increases significantly and eventually diverges, signalling the emergence of pronounced critical slowing down in the kinetic evolution.

In Fig.~\ref{f2a}, we display the temperature profile together with the corresponding spinodal lines that arise when the angular momentum lies below its critical value. The associated free energy landscapes at the spinodal temperatures are shown in Fig.~\ref{f2b}. The blue and red curves correspond respectively to the left and right spinodal temperatures, while the black curve represents the landscape at an intermediate temperature between the two spinodal points. One can clearly observe the gradual deformation of the landscape structure as the system evolves between the two spinodal regimes.

\begin{figure}[t!]
    \centering
    \begin{subfigure}[b]{0.45\textwidth}
        \centering
        \includegraphics[width=\textwidth]{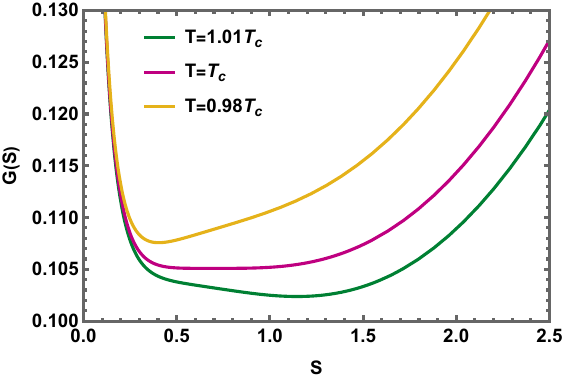}
        \caption{}
        \label{f3a}
    \end{subfigure}
    \hspace{0.8cm}
    \begin{subfigure}[b]{0.45\textwidth}
        \centering
        \includegraphics[width=\textwidth]{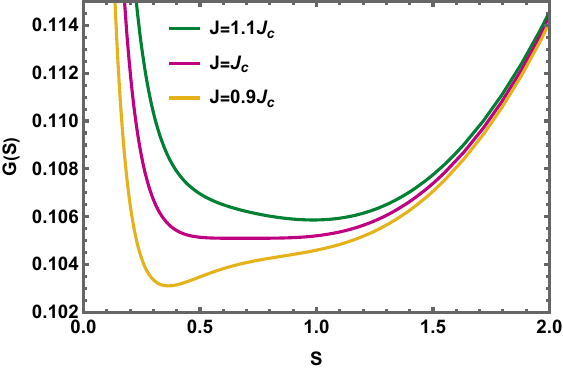}
        \caption{}
        \label{f3b}
    \end{subfigure}

    \caption{Free energy landscapes for varying ensemble temperature (left) and angular momentum (right).}
    \label{f3}
\end{figure}

Figure~\ref{f3} illustrates the behaviour of the free energy landscape for varying ensemble temperature $T$ (Fig.~\ref{f3a}) and varying angular momentum $J$ (Fig.~\ref{f3b}). In both cases, the landscape becomes significantly flatter as the system approaches the critical point, represented by the magenta curve corresponding to the critical temperature or critical angular momentum. This flattening signals the suppression of the effective restoring force near equilibrium and plays a crucial role in the emergence of critical slowing down. Away from criticality, however, the free energy regains a well-defined single-well structure when the temperature or angular momentum is chosen slightly above or below their corresponding critical values.

\subsection{Numerical Results}
We now investigate the stochastic dynamics numerically by simulating the Langevin equation governing the evolution of the order parameter on the free energy landscape. From the resulting stochastic trajectories, we compute the autocorrelation function and extract the corresponding autocorrelation time by fitting it to an exponential decay profile. In addition to the autocorrelation time, we also analyse the variance of the trajectories together with the smallest nonzero eigenvalue of the associated Fokker-Planck equation in order to further characterize the relaxation dynamics of the system. While the variance measures the strength of fluctuations around equilibrium configurations, the smallest eigenvalue of the Fokker-Planck operator provides an estimate of the characteristic relaxation timescale. For convenience, we set $\zeta = 1$ throughout the numerical analysis. The numerical procedures employed in this work are summarized in the Appendix \ref{appendix A}.

\begin{figure}[t!]
\centering
\includegraphics[width=0.97\textwidth]{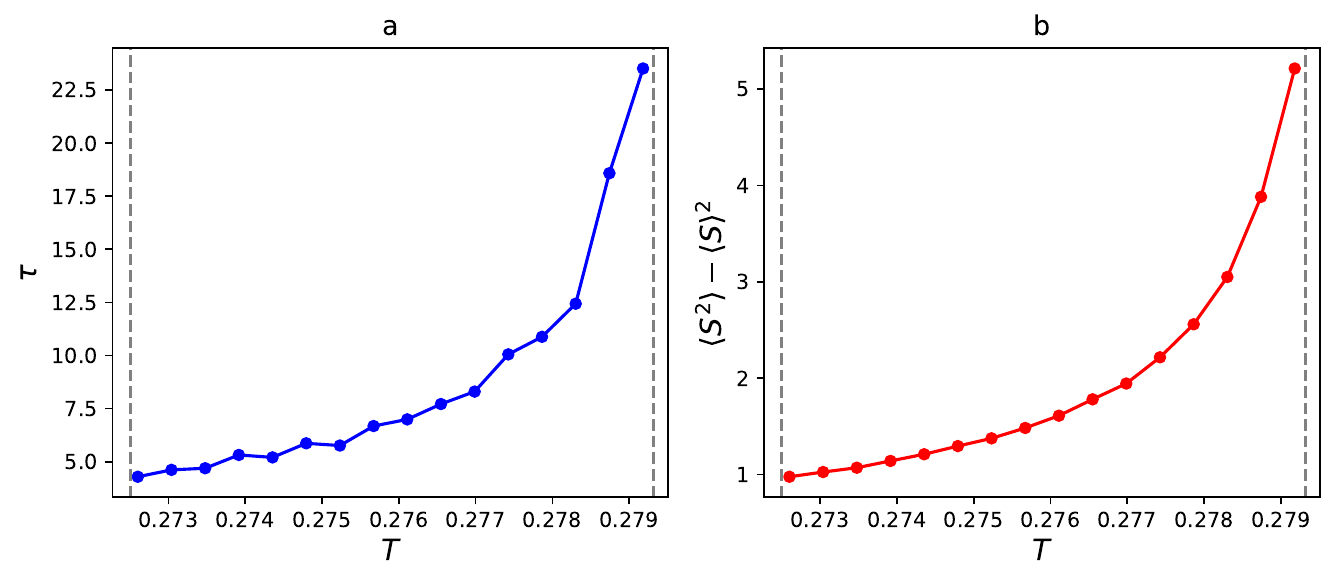}
\caption{Correlation time $\tau$ and variance $\langle S^2\rangle-\langle S\rangle^2$ of trajectories as a function of $T$ near small black hole.}
\label{f4}
\end{figure}

We first investigate the stochastic dynamics in the vicinity of the spinodal points. As the temperature is increased from the left spinodal temperature toward the right spinodal point, the free energy well corresponding to the small black hole phase progressively becomes shallower and eventually disappears. This behaviour is illustrated in Fig.~\ref{f2b}, where the blue and red curves correspond respectively to the left and right spinodal temperatures. The resulting flattening of the free energy landscape weakens the effective restoring force near equilibrium, leading to increasingly slow relaxation dynamics. Consequently, both the autocorrelation time and the variance of the stochastic trajectories associated with the small black hole phase increase significantly with temperature, as shown in the left panels of Fig.~\ref{f4}, providing a clear signature of critical slowing down.

Conversely, when the temperature is lowered from the right spinodal temperature toward the left spinodal point (from the red curve to the blue curve in Fig~\ref{f2b}), the free energy well associated with the large black hole phase gradually becomes flatter and eventually disappears.
\begin{figure}[h!]
\centering
\includegraphics[width=0.97\textwidth]{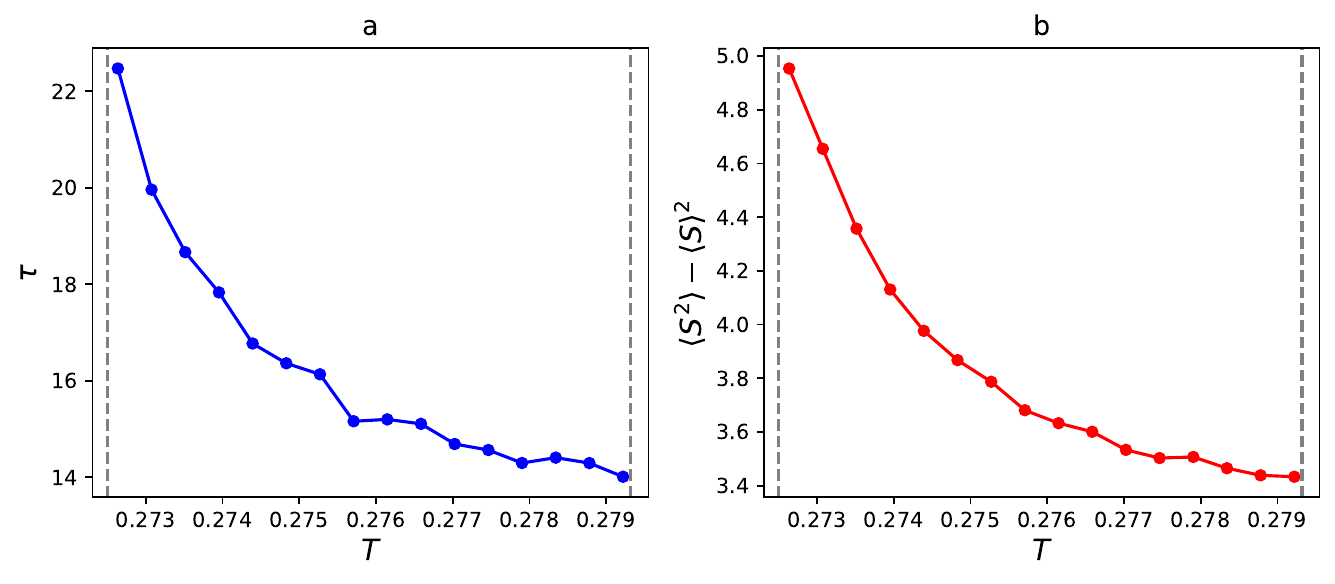}
\caption{Correlation time $\tau$ and variance $\langle S^2\rangle-\langle S\rangle^2$ of trajectories as a function of $T$ near large black hole.}
\label{f5}
\end{figure} As in the small black hole phase, the reduction of the local curvature near equilibrium leads to slower relaxation dynamics. Consequently, both the autocorrelation time and the variance of the stochastic trajectories corresponding to the large black hole phase increase as the temperature decreases, as illustrated in Fig.~\ref{f5}, again indicating the presence of critical slowing down.
\begin{figure}[t!]
\centering
\includegraphics[width=0.97\textwidth]{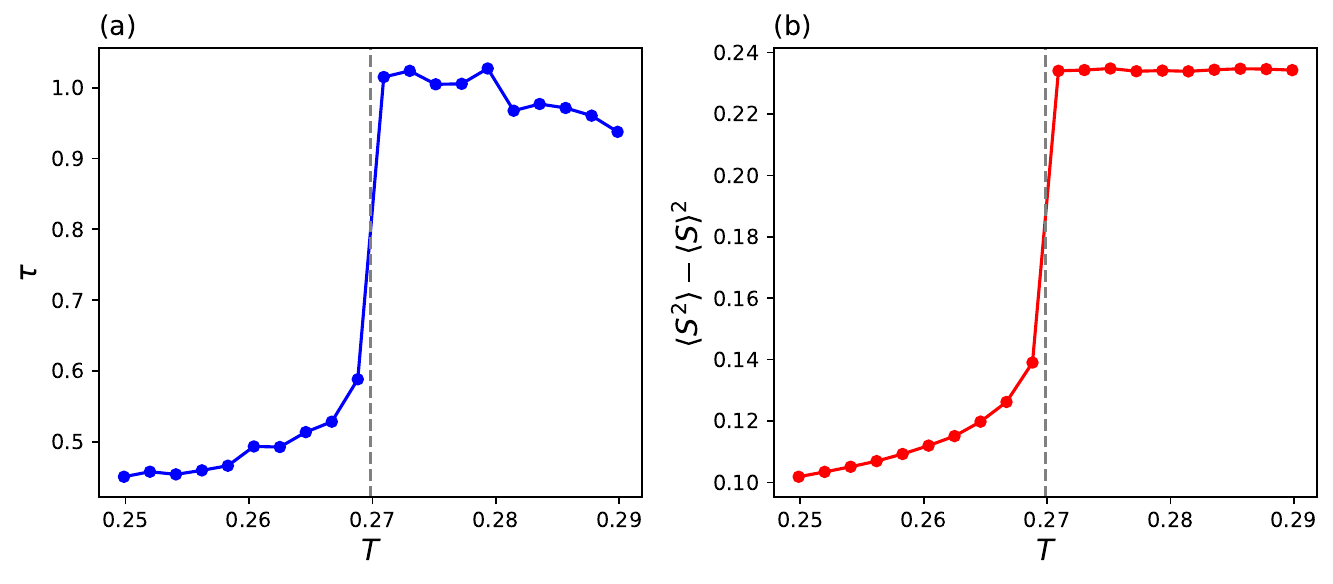}
\caption{Correlation time $\tau$ and variance $\langle S^2\rangle-\langle S\rangle^2$ of trajectories as a function of $T$. The dashed line is represents the critical temperature.}
\label{f6}
\end{figure}

We next investigate the stochastic dynamics in the vicinity of the critical point. To this end, we consider two distinct thermodynamic paths. In the first case, the angular momentum is fixed at its critical value, $J=J_c$, while the temperature is varied in the neighbourhood of $T_c$. In the second case, the temperature is fixed at $T=T_c$, while the angular momentum is varied around its critical value $J_c$. 
\begin{figure}[h!]
\centering
\includegraphics[width=0.97\textwidth]{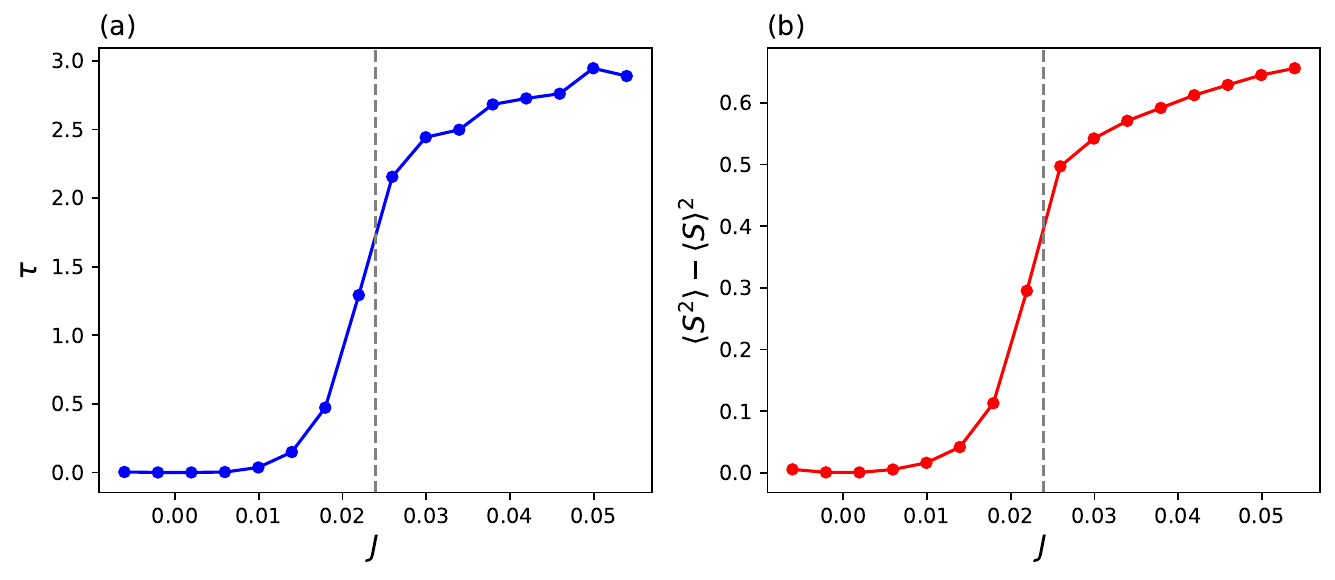}
\caption{Correlation time $\tau$ and variance $\langle S^2\rangle-\langle S\rangle^2$ of trajectories as a function of $J$. The dashed line is represents the critical angular momentum.}
\label{f7}
\end{figure}
The corresponding behaviour of the autocorrelation time and variance along these trajectories is shown in Figs.~\ref{f6} and \ref{f7}.

As the system approaches criticality along either path, both the autocorrelation time and the variance exhibit a pronounced increase, reaching their maximum values near the critical point. This behaviour reflects the progressive flattening of the free energy landscape at criticality, as illustrated in Fig.~\ref{f3}. Away from the critical point, both the autocorrelation time and the variance remain approximately constant with changing temperature and angular momentum, as shown in Fig.~\ref{f6} and \ref{f7}.

\begin{figure}[h!]
    \centering
    \begin{subfigure}[b]{0.45\textwidth}
        \centering
        \includegraphics[width=\textwidth]{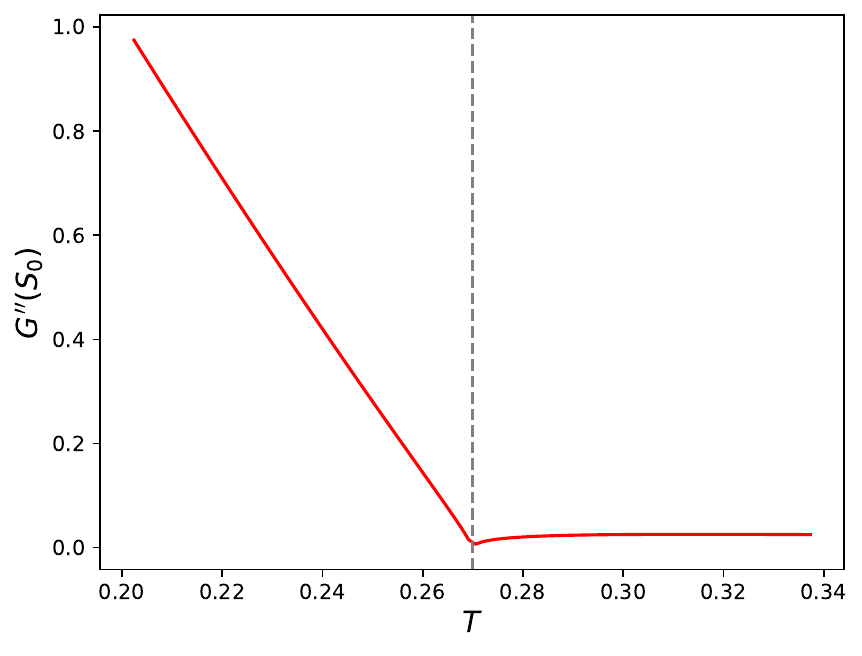}
        \caption{}
        \label{f8a}
    \end{subfigure}
    \hspace{0.8cm}
    \begin{subfigure}[b]{0.45\textwidth}
        \centering
        \includegraphics[width=\textwidth]{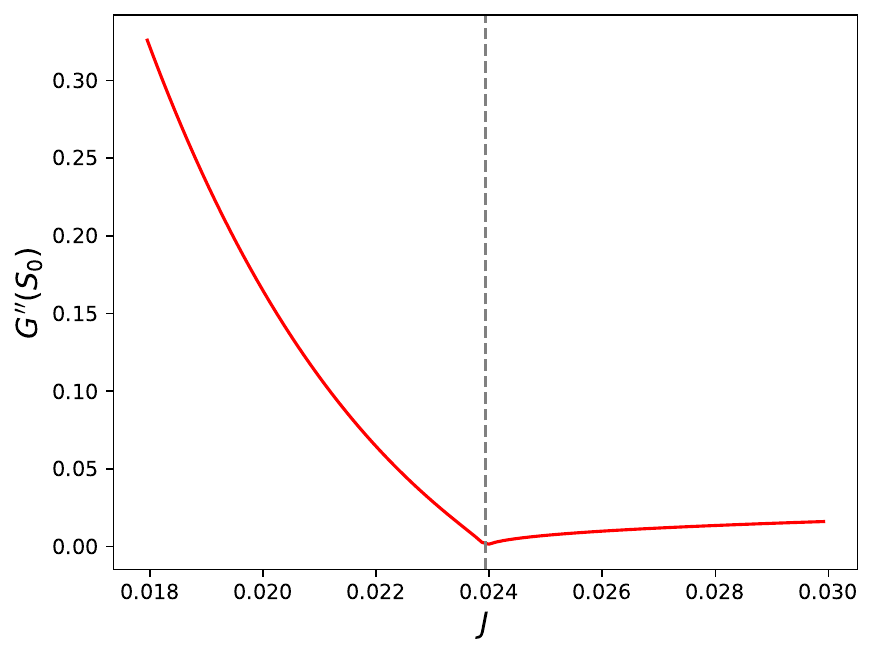}
        \caption{}
        \label{f8b}
    \end{subfigure}

    \caption{Local curvature of free energy landscapes with varying ensemble temperature (left) and angular momentum (right).}
    \label{f8}
\end{figure}

To further substantiate this behaviour, we analyse the local curvature of the free energy landscape by examining $G''(S_0)$ as a function of temperature and angular momentum, shown in Figs.~\ref{f8a} and \ref{f8b}. The results demonstrate that the curvature decreases steadily as the critical point is approached and attains its minimum value near criticality, consistent with the flattening of the free energy landscape. Away from the critical point, the curvature remains approximately constant, which is reflected in the nearly unchanged behaviour of the autocorrelation time and variance observed in Fig.~\ref{f6} and \ref{f7}.

\begin{figure}[h!]
    \centering
    \begin{subfigure}[b]{0.466\textwidth}
        \centering
        \includegraphics[width=\textwidth]{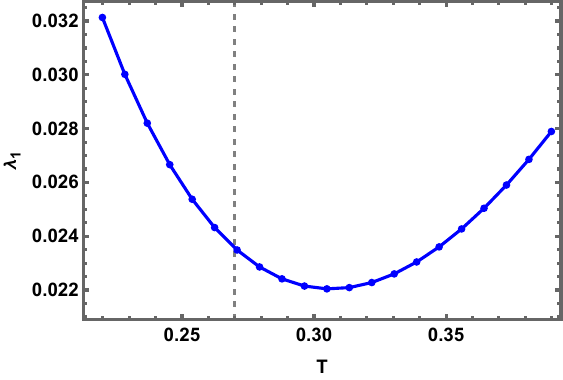}
        \caption{}
        \label{f9a}
    \end{subfigure}
    \hspace{0.8cm}
    \begin{subfigure}[b]{0.45\textwidth}
        \centering
        \includegraphics[width=\textwidth]{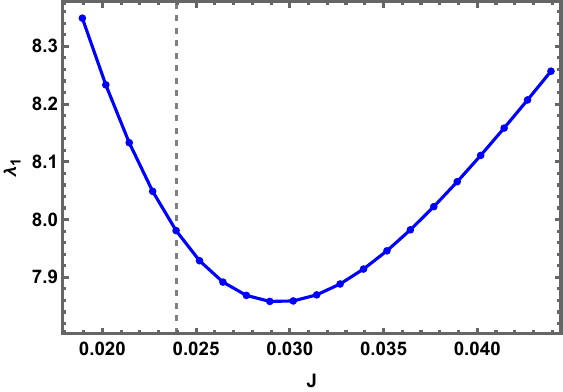}
        \caption{}
        \label{f9b}
    \end{subfigure}

    \caption{The variation of the smallest non-zero eigenvalue of the Fokker-Planck equation with (a) temperature and (b) pressure. The grey dotted line represents the critical value of the temperature and pressure.}
    \label{f9}
\end{figure}

We further investigate the relaxation dynamics through the smallest nonzero eigenvalue of the associated Fokker-Planck equation. As shown in Fig.~\ref{f9}, the eigenvalue exhibits a pronounced suppression in the vicinity of the critical point, consistent with the significant enhancement of the autocorrelation time and trajectory variance observed earlier. This behaviour provides additional evidence for the emergence of critical slowing down near criticality. The numerical procedure used to obtain the eigenvalues from the Fokker-Planck equation is summarized in the Appendix \ref{appendix A}.

Interestingly, the slowest relaxation dynamics does not occur exactly at the critical point. Along both the paths, with fixed angular momentum or fixed temperature, the strongest slowing down appears slightly below the critical temperature or angular momentum. A similar shift of the lowest eigenvalue away from the exact critical point was also reported in the RN-AdS case.

\section{Scaling Law and Universality}\label{sec4}
\subsection{Numerical Extraction of Delta}
The results of the previous section demonstrate a pronounced enhancement of the relaxation time near criticality, signalling the emergence of critical slowing down in black hole phase transitions. An important question is whether this behaviour exhibits a universal scaling structure. To address this, we investigate the scaling behaviour of the autocorrelation time near the critical point for several black hole systems. We begin with the RN-AdS case, for which the dynamical scaling exponent has not been explicitly analysed in earlier studies, and subsequently extend the analysis to Kerr-AdS and Bardeen black holes in order to examine the possible universality of the scaling behaviour.

The autocorrelation time is expected to obey a power-law scaling of the form \cite{slowing}
\begin{equation}\label{eq4.1}
\tau \sim |\epsilon|^{-\Delta},
\end{equation}
where $\epsilon$ is termed as the reduced temperature and $\Delta$ is the dynamical critical exponent.

We can define
\begin{equation}\label{eq4.2}
\epsilon = \frac{T-T_c}{T_c},\quad \epsilon_P=\frac{P-P_c}{P_c},\quad \epsilon_J=\frac{J-J_c}{J_c}
\end{equation}
along different thermodynamic paths.

\begin{figure}[h!]
    \centering
 \begin{subfigure}{\textwidth}
        \centering
        \includegraphics[width=\linewidth]{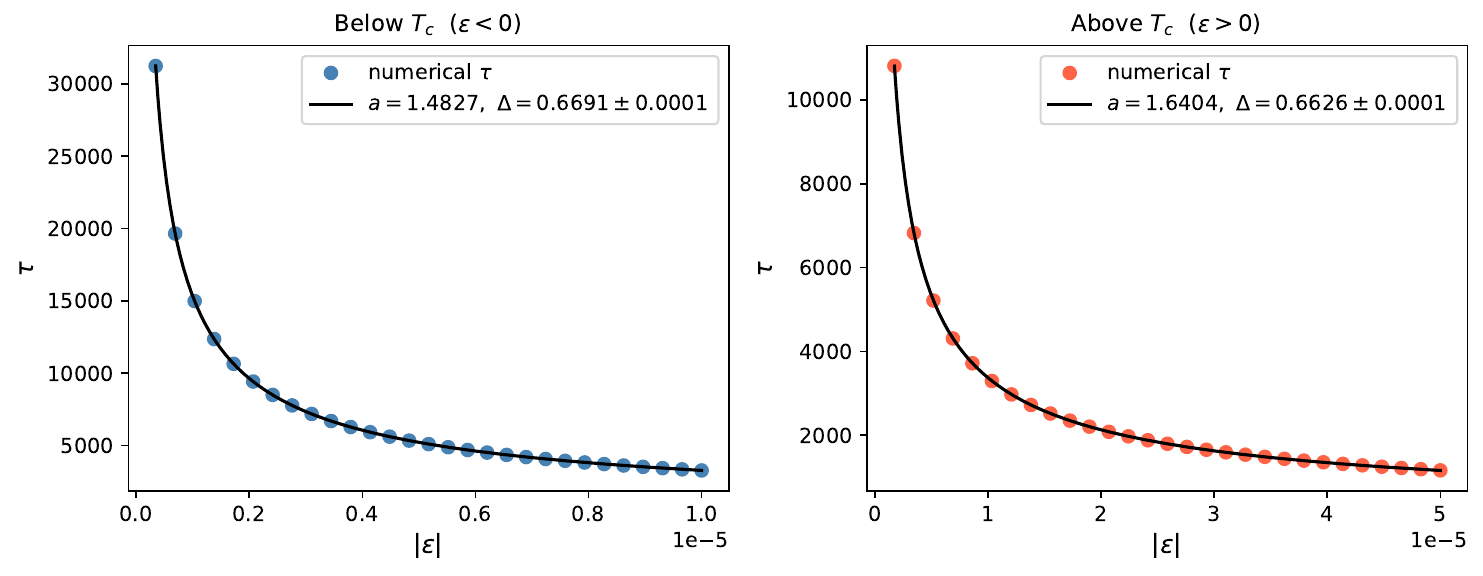}
        \caption{Scaling behaviour of the autocorrelation time near criticality along temperature.}
        \label{f10a}
    \end{subfigure}

    \vspace{0.5cm}

    \begin{subfigure}{\textwidth}
        \centering
        \includegraphics[width=\linewidth]{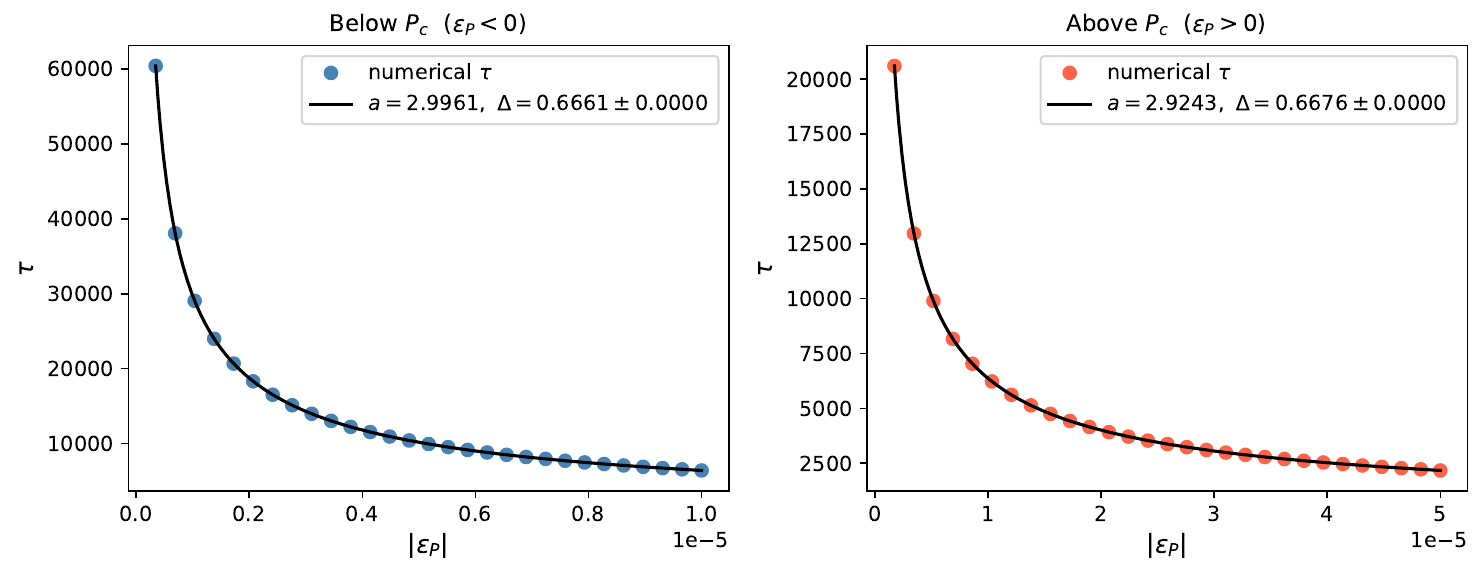}
        \caption{Scaling behaviour of the autocorrelation time near criticality along pressure.}
        \label{f10b}
    \end{subfigure}

    \caption{Numerically obtained autocorrelation time $\tau$ for the RN-AdS black hole together with the corresponding fits to the scaling form $\tau=a|\epsilon|^{-\Delta}$ near the critical point.}
    \label{f10}
\end{figure}

To determine the dynamical critical exponent, we fit the numerically obtained autocorrelation time near the critical point to the scaling form
\begin{equation}
\tau = a |\epsilon|^{-\Delta}
\end{equation}
where $a$ and $\Delta$ are treated as fitting parameters.

In Fig.~\ref{f10}, we present the numerically obtained autocorrelation time together with the corresponding fits to the scaling form $\tau = a |\epsilon|^{-\Delta}$ near the criticality. Two thermodynamic paths are considered: varying the temperature at fixed pressure (Fig.~\ref{f10a}) and varying the pressure at fixed temperature (Fig.~\ref{f10b}). In all cases, the extracted dynamical critical exponent is found to be approximately $\Delta \simeq 0.66$, indicating a robust scaling behaviour near criticality.

\begin{figure}[ht]
    \centering
 \begin{subfigure}{\textwidth}
        \centering
        \includegraphics[width=\linewidth]{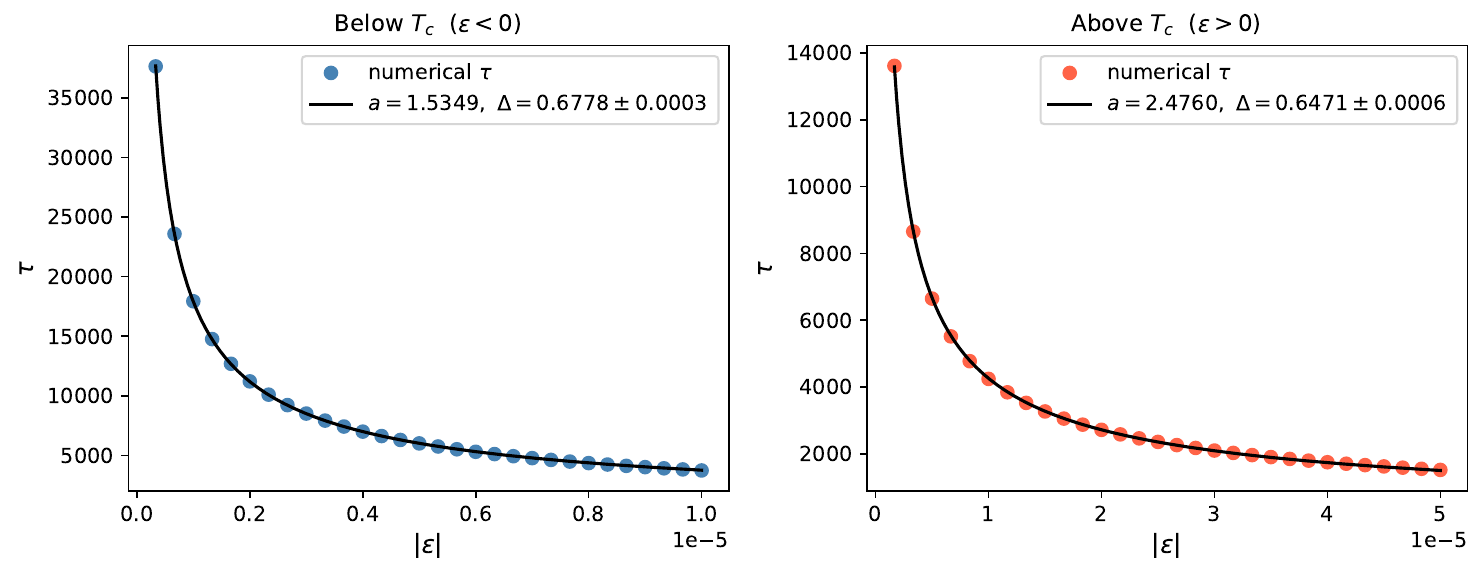}
        \caption{Scaling behaviour of the autocorrelation time near criticality along temperature.}
        \label{f11a}
    \end{subfigure}

    \vspace{0.5cm}

    \begin{subfigure}{\textwidth}
        \centering
        \includegraphics[width=\linewidth]{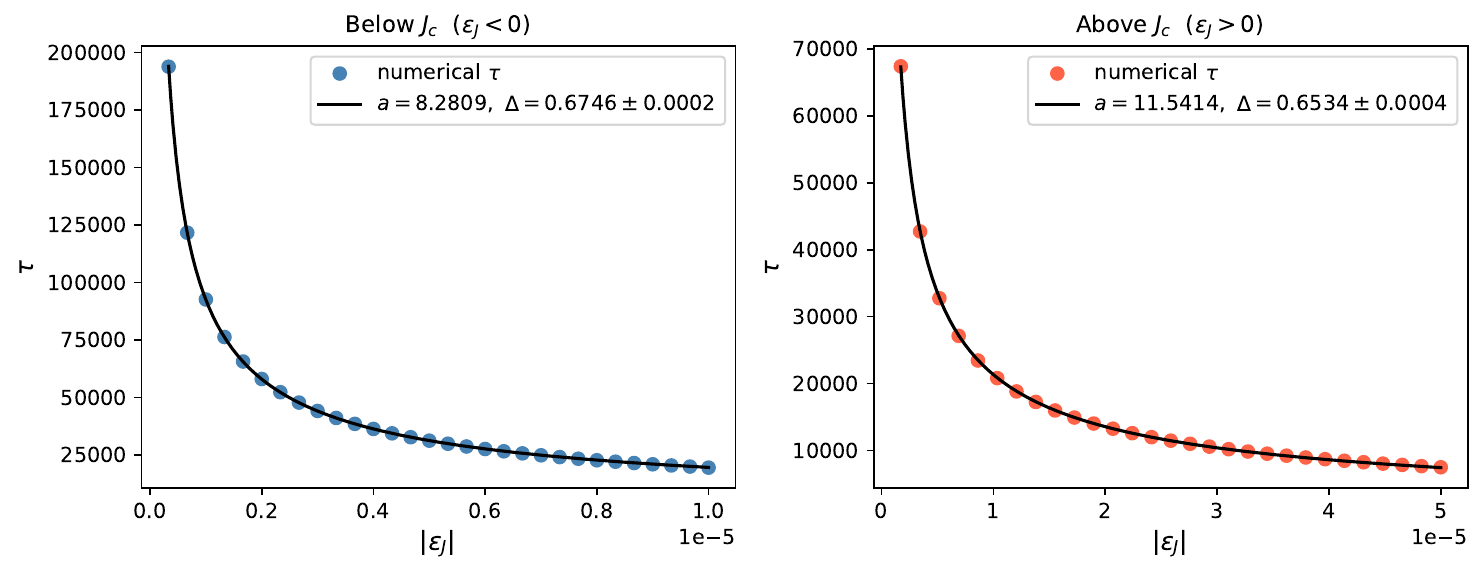}
        \caption{Scaling behaviour of the autocorrelation time near criticality along angular momentum.}
        \label{f11b}
    \end{subfigure}

    \caption{Numerically obtained autocorrelation time $\tau$ for the Kerr-AdS black hole together with the corresponding fits to the scaling form $\tau=a|\epsilon|^{-\Delta}$ near the critical point.}
    \label{f11}
\end{figure}

The prefactor $a$, however, is observed to depend both on the thermodynamic path and on the direction from which the critical point is approached, namely $\epsilon<0$ or $\epsilon>0$. Furthermore, as $\epsilon \rightarrow 0$, the autocorrelation time increases sharply, consistent with the expected divergence of the relaxation timescale at the critical point. The extracted exponent is remarkably close to the analytical value $\Delta = 2/3$. A derivation of this value from the critical expansion of the free energy landscape is presented in the Appendix \ref{appendix B}.

We next turn to the Kerr-AdS system and analyse the scaling behaviour of the autocorrelation time using the same fitting relation, $\tau = a|\epsilon|^{-\Delta}$. The corresponding fits and the extracted values of the fitting parameters are presented in Fig.~\ref{f11}. As in the RN-AdS case, the dynamical critical exponent is consistently found to be approximately $\Delta \simeq 0.66 \approx 2/3$ for both thermodynamic paths, namely varying the temperature at fixed angular momentum and varying the angular momentum at fixed temperature, as well as for both directions of approach to the critical point. 

These results indicate that the dynamical scaling exponent remains unchanged in both charged and rotating black hole systems, suggesting the existence of an underlying universal scaling behaviour.

In order to further confirm our result about this universal scaling behaviour, we consider a regular black hole system - the Bardeen black hole. Bardeen black holes constitute one of the earliest examples of regular black hole solutions that satisfy the weak energy condition~\cite{bardeen}. In contrast to conventional singular black holes such as the Schwarzschild or Kerr solutions, they possess a nonsingular core, leading to a regular spacetime geometry at the centre. The Bardeen black hole has the following generalised free energy
\begin{equation}
G(r_+)=\frac{\left(g^2+r_+^2\right){}^{3/2} \left(8 \pi  P r_+^2+3\right)}{6 r_+^2}-T\pi  r_+^2 
\end{equation}

Since the analysis exactly similar to the RN-AdS and Kerr-AdS cases, we do not repeat the full stochastic Langevin simulation for the Bardeen black hole. Instead, we rely on the analytical argument that the flattening of the generalized free energy landscape near criticality should likewise lead to the emergence of critical slowing down. 

\begin{figure}[h]
    \centering
    \begin{subfigure}[b]{0.45\textwidth}
        \centering
        \includegraphics[width=\textwidth]{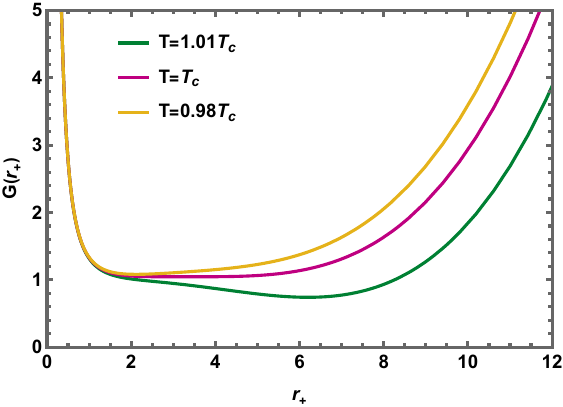}
        \caption{}
        \label{f12a}
    \end{subfigure}
    \hspace{0.8cm}
    \begin{subfigure}[b]{0.45\textwidth}
        \centering
        \includegraphics[width=\textwidth]{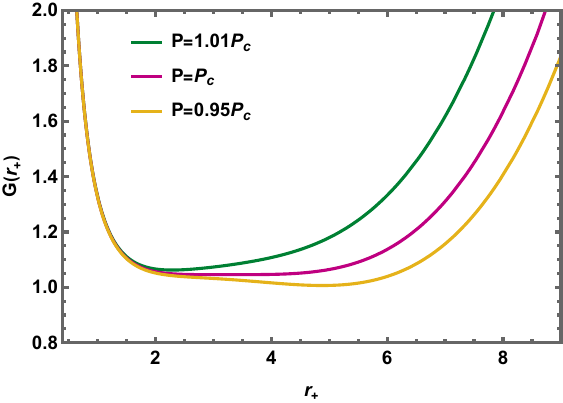}
        \caption{}
        \label{f12b}
    \end{subfigure}

    \caption{Free energy landscapes for varying ensemble temperature (left) and pressure (right).}
    \label{f12}
\end{figure}

The corresponding free energy landscapes are shown in Fig.~\ref{f12}. In both cases, the landscape progressively flattens as the critical values of temperature and pressure are approached, with the magenta curves in Figs.~\ref{f12a} and \ref{f12b} representing the critical configurations. This behaviour indicates the vanishing of the local curvature of the free energy landscape near equilibrium and therefore suggests the onset of slow relaxation dynamics, analogous to that observed in the RN-AdS and Kerr-AdS systems.

\begin{figure}[H]
    \centering
 \begin{subfigure}{\textwidth}
        \centering
        \includegraphics[width=\linewidth]{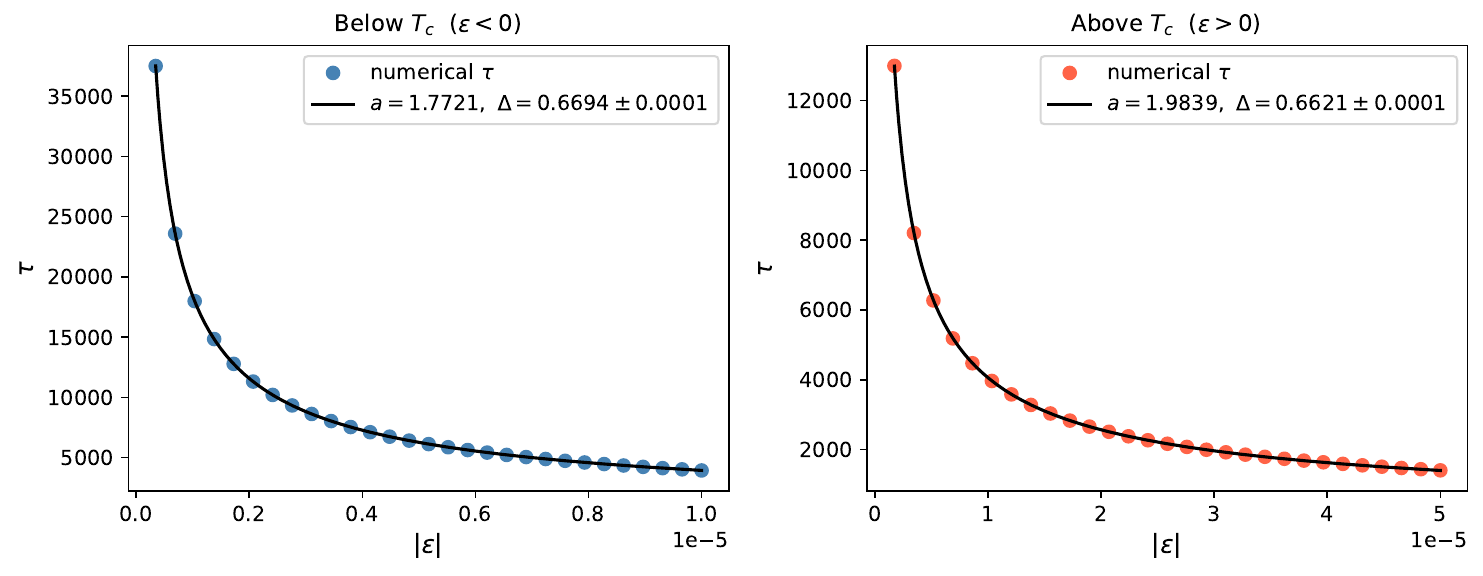}
        \caption{Scaling behaviour of the autocorrelation time near criticality along temperature.}
        \label{f13a}
    \end{subfigure}

    \vspace{0.5cm}

    \begin{subfigure}{\textwidth}
        \centering
        \includegraphics[width=\linewidth]{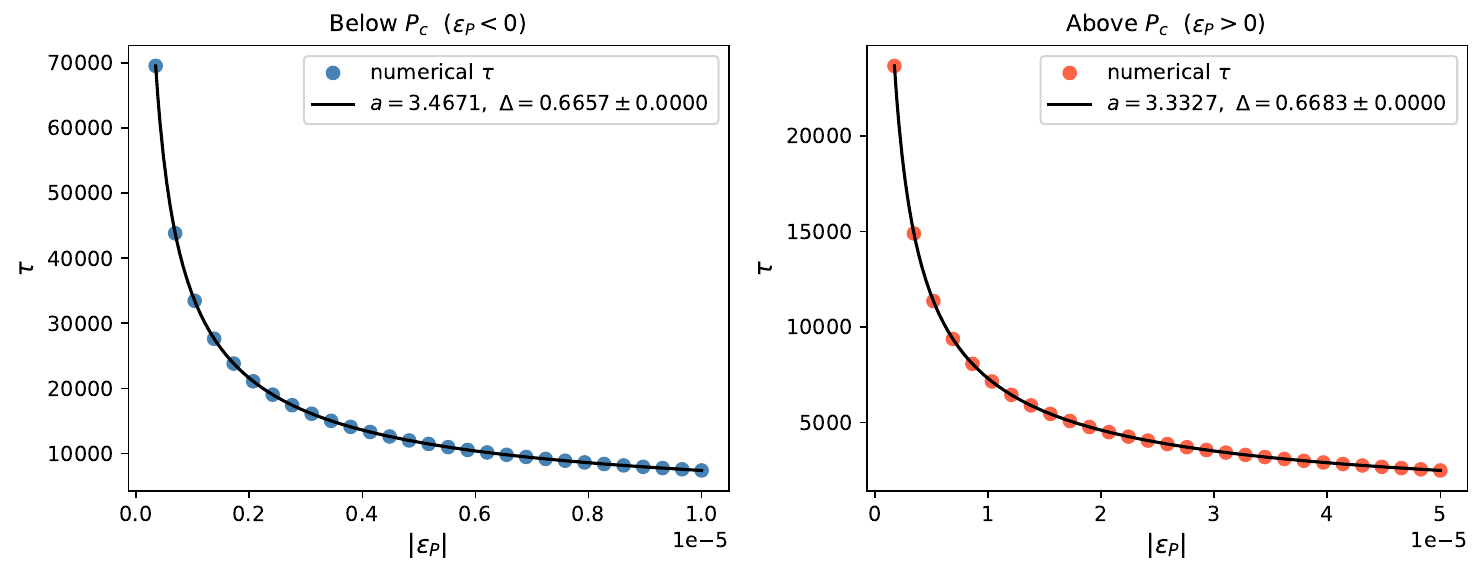}
        \caption{Scaling behaviour of the autocorrelation time near criticality along pressure.}
        \label{f13b}
    \end{subfigure}

    \caption{Numerically obtained autocorrelation time $\tau$ for the Bardeen black hole together with the corresponding fits to the scaling form $\tau=a|\epsilon|^{-\Delta}$ near the critical point.}
    \label{f13}
\end{figure}

We therefore proceed directly to the analysis of the scaling behaviour near criticality. Employing the same fitting procedure used in the RN-AdS and Kerr-AdS cases, we extract the dynamical critical exponent $\Delta$ by fitting the autocorrelation time to the scaling form
$\tau = a|\epsilon|^{-\Delta}$. The corresponding numerical results are presented in Fig.~\ref{f13}.

For the Bardeen black hole as well, the dynamical critical exponent is consistently found to be approximately $\Delta \simeq 0.66 \approx 2/3$ along both thermodynamic paths, namely varying the temperature and varying the pressure, as well as for both directions of approach to the critical point. The persistence of the same scaling exponent across charged, rotating, and regular black hole systems strongly suggests the existence of a universal dynamical behaviour near criticality.
\subsection{Delta from the Fokker-Planck spectrum}
The relaxation dynamics near criticality may also be characterized through the spectrum of the associated Fokker--Planck operator. The probability distribution $\rho(x,t)$ for fluctuations of the order parameter evolves according to \cite{Li,Li2}
\begin{equation}
\frac{\partial P(x,t)}{\partial t}=D \frac{\partial}{\partial x}\left\{e^{-\beta G(x)}\frac{\partial}{\partial x}\left[e^{\beta G(x)} P(x,t)\right]\right\},
\end{equation}
where $D=k_BT/\zeta$ is the diffusion coefficient and $G(x)$ denotes the generalized free energy landscape.
Near equilibrium, the slowest relaxation process is governed by the smallest nonvanishing eigenvalue $\lambda_1$ of the Fokker-Planck operator, with the corresponding relaxation time given by $\tau \sim \frac{1}{\lambda_1}$.
As the critical point is approached, the curvature of the free energy landscape decreases and eventually vanishes, leading to the suppression of the spectral gap $\lambda_1$. Consequently, the relaxation time diverges, signalling the onset of critical slowing down. Near criticality, the relaxation time scales as $\tau \sim |\epsilon|^{-\Delta}$, consequently, the smallest nonzero eigenvalue should obey the scaling relation $\lambda_1 \sim |\epsilon|^{\gamma}$ \cite{slowing}. The exponents are extracted from linear fits in the corresponding log-log plots.

\begin{figure}[h]
\centering
\includegraphics[width=0.97\textwidth]{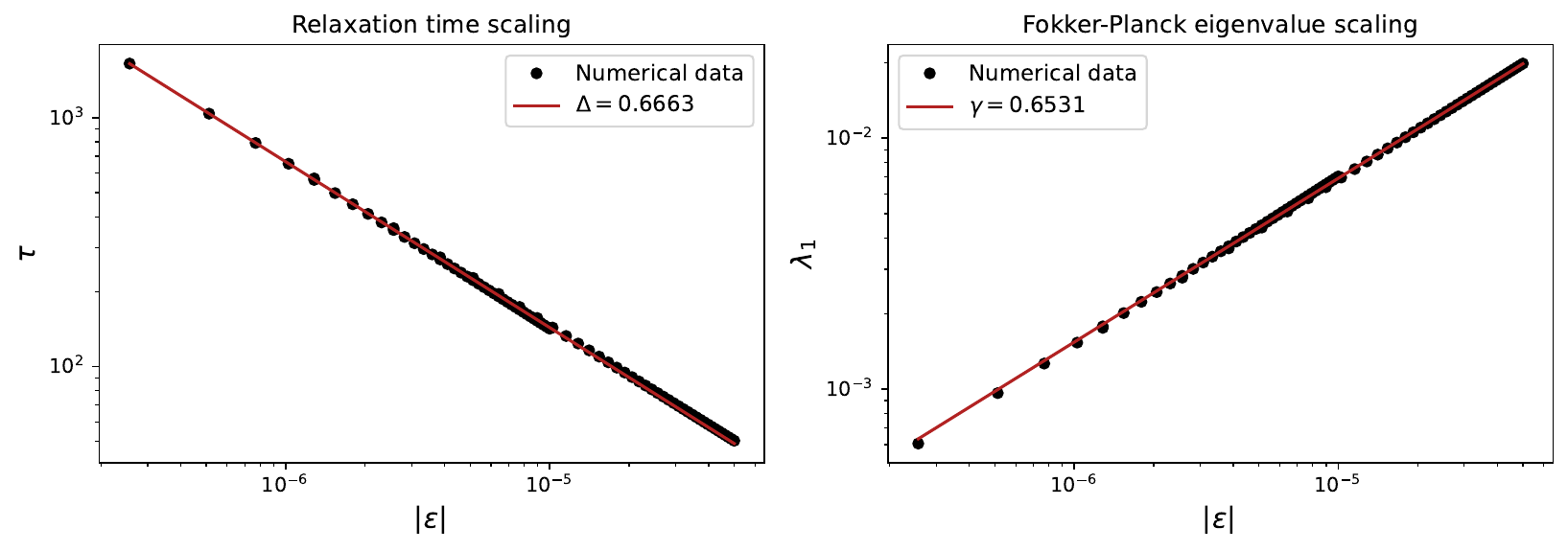}
\caption{Log-log plots showing the scaling behaviour of the relaxation time $\tau$ and the smallest nonzero Fokker-Planck eigenvalue $\lambda_1$ near criticality for RN-AdS black hole.}
\label{f14}
\end{figure}

The numerical results for the relaxation time and the smallest nonzero Fokker-Planck eigenvalue in the RN-AdS black hole are presented in Fig.~\ref{f14}. In both cases, the extracted scaling exponents are found to be approximately $\Delta \simeq \gamma \simeq 0.66 \approx 2/3$, consistent with the expected inverse relation between the relaxation time and the spectral gap. This provides an independent confirmation of the scaling exponent previously obtained from the autocorrelation time analysis. 

\begin{figure}[h]
    \centering
 \begin{subfigure}{\textwidth}
        \centering
        \includegraphics[width=\linewidth]{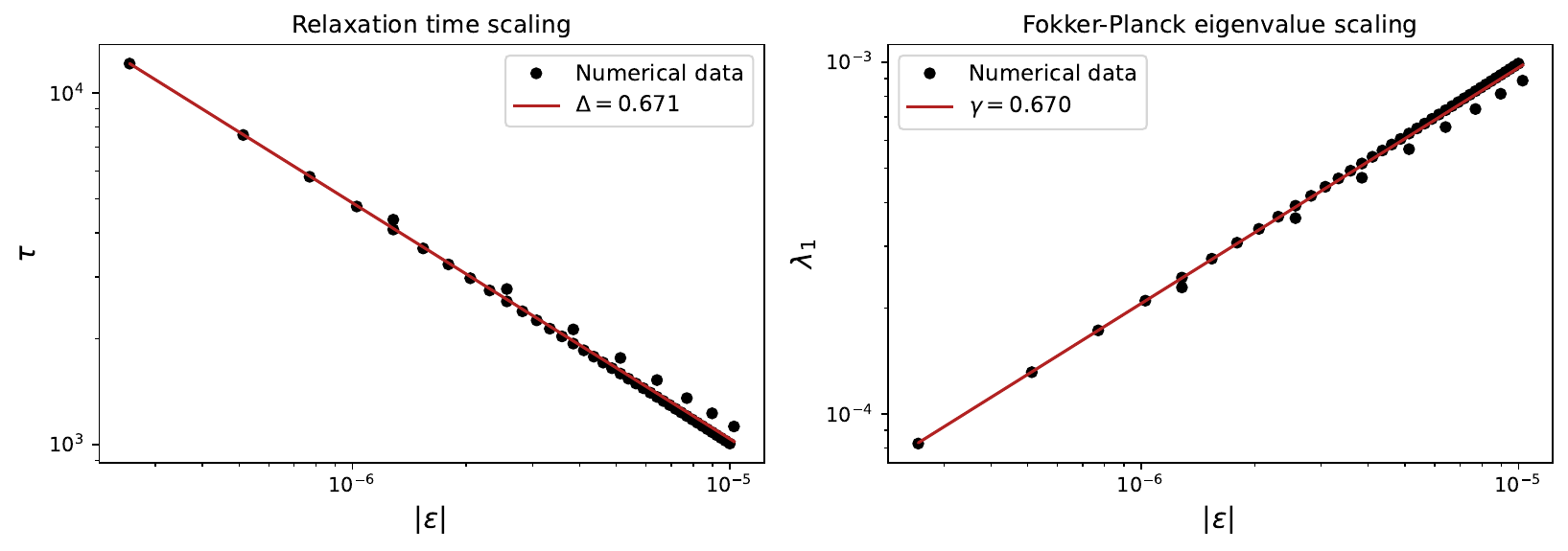}
        \caption{Scaling behaviour of the relaxation time and the smallest nonzero Fokker-Planck eigenvalue for the Kerr-AdS black hole.}
        \label{f15a}
    \end{subfigure}

    \vspace{0.5cm}

    \begin{subfigure}{\textwidth}
        \centering
        \includegraphics[width=\linewidth]{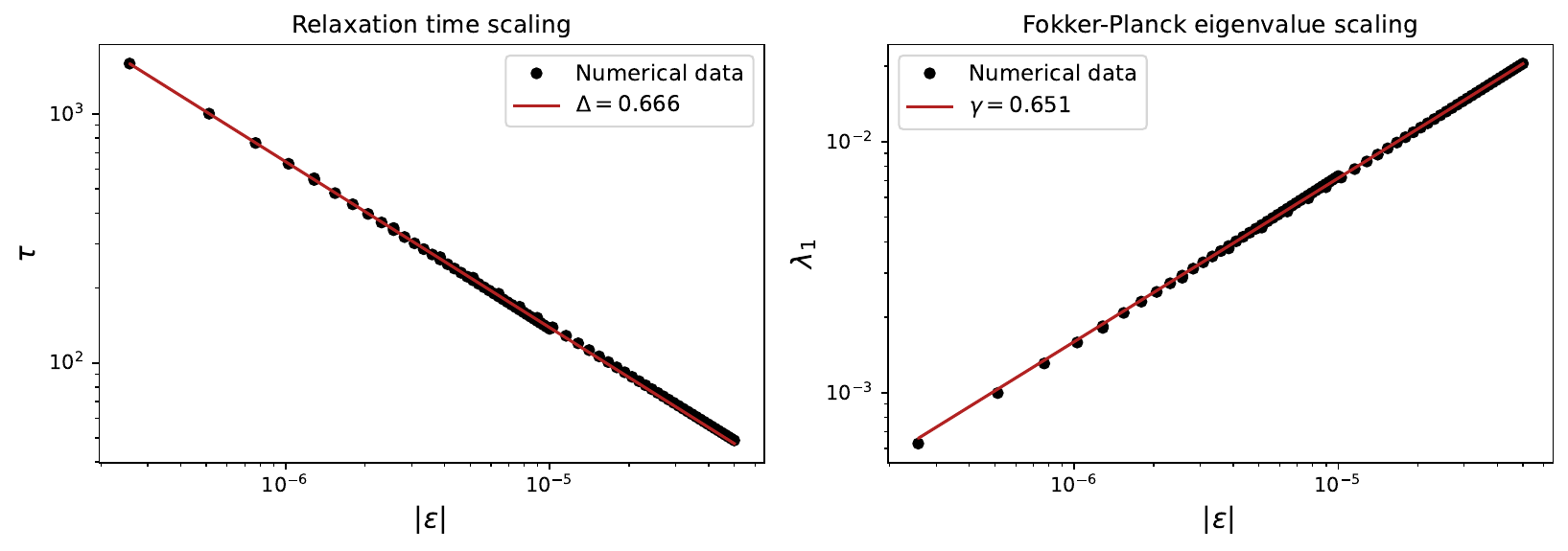}
        \caption{Scaling behaviour of the relaxation time and the smallest nonzero Fokker-Planck eigenvalue for the Bardeen black hole.}
        \label{f15b}
    \end{subfigure}

    \caption{Log-log plots showing the scaling behaviour of the relaxation time $\tau$ and the smallest nonzero Fokker-Planck eigenvalue $\lambda_1$ near criticality.}
    \label{f15}
\end{figure}

We further extend the analysis to the Kerr-AdS and Bardeen black hole systems, with the corresponding numerical results presented in Fig.~\ref{f15}.

In both cases, the scaling exponents extracted from the relaxation time and the smallest nonzero Fokker--Planck eigenvalue are consistently found to satisfy $\Delta \simeq \gamma \simeq 0.66 \approx \frac{2}{3}$. The persistence of the same scaling exponent across charged, rotating, and regular black hole backgrounds strongly suggests the existence of a universal dynamical scaling behaviour near criticality. These results indicate that the emergence of critical slowing down is governed primarily by the underlying structure of the free energy landscape, rather than by the specific details of the black hole geometry.

\section{Conclusion and Discussion}\label{sec5}

In this work, we investigated the dynamical critical behaviour associated with black hole phase transitions in anti-de Sitter spacetime within the framework of stochastic free energy landscape dynamics. Extending earlier analyses performed for RN-AdS black holes, we studied the emergence of critical slowing down in Kerr-AdS black holes by treating the black hole entropy as the relevant dynamical order parameter and analysing its stochastic evolution near criticality through the Langevin and Fokker-Planck formalisms. Our results demonstrate that the relaxation dynamics of the system becomes significantly slower as the critical point is approached, reflected in the enhancement of the autocorrelation time, the growth of fluctuations, and the suppression of the smallest nonzero eigenvalue of the associated Fokker-Planck operator.

We showed that the origin of this slowing down can be understood from the progressive flattening of the generalized free energy landscape near the critical point. As the local curvature of the free energy decreases, the effective restoring force governing the relaxation dynamics weakens, leading to increasingly long-lived fluctuations. Similar behaviour was also observed near the spinodal points, where the free energy landscape develops an analogous flattening structure and the relaxation timescale correspondingly increases.

A central result of the present work is the emergence of a robust dynamical scaling law near criticality. By analysing the scaling behaviour of the autocorrelation time along different thermodynamic paths, including variations in temperature, pressure, and angular momentum, we found that the relaxation time follows the power-law behaviour $\tau \sim |\epsilon|^{-2/3}$, where $\epsilon$ denotes the reduced thermodynamic parameter measuring the distance from the critical point. The same exponent was independently recovered from the scaling behaviour of the smallest nonzero Fokker--Planck eigenvalue. Importantly, this scaling structure was found not only for RN-AdS and Kerr-AdS black holes, but also for the regular Bardeen black hole system. The persistence of the exponent $\Delta \simeq 2/3$ across charged, rotating, and regular black hole backgrounds strongly suggests the existence of a universal dynamical behaviour underlying black hole critical phenomena.

Our results suggest that the emergence of critical slowing down is mainly controlled by the flattening of the free energy landscape near criticality, rather than by the specific nature of the black hole system itself. Since the same scaling behaviour is observed for RN-AdS, Kerr-AdS, and Bardeen black holes, the dynamical critical behaviour appears to exhibit a universal character.

The present work opens several possible directions for future investigation. It would be interesting to extend the analysis to higher-dimensional black holes, modified gravity theories, and multicharge or multicritical systems in order to further examine the universality of the dynamical scaling exponent. Another important direction would be to explore possible connections between the stochastic critical dynamics discussed here and other probes of black hole phase structure, including thermodynamic geometry, quasinormal modes, Lyapunov dynamics, and holographic nonequilibrium phenomena. We hope that the present study provides further insight into the nonequilibrium aspects of black hole thermodynamics and the dynamical nature of gravitational phase transitions.

\acknowledgments
The authors would like to thank Saumen Acharjee for valuable discussions and insightful comments.
\appendix
\section{Numerical Implementation}\label{appendix A}
In this appendix we explain the numerical techniques used to integrate the Langevin equation and the Fokker-Planck equation

In this appendix, we briefly summarize the numerical methods employed for integrating the Langevin equation and computing the relaxation dynamics near criticality.

The stochastic evolution of the order parameter $S$ is governed by the overdamped Langevin equation
\begin{equation}
\frac{dS}{dt}=-\frac{1}{\zeta}\frac{\partial G(S)}{\partial S}+\eta(t),
\end{equation}
where $G(S)$ denotes the generalized free energy landscape, $\zeta$ is the damping coefficient, and $\eta(t)$ represents Gaussian white noise satisfying
\begin{equation}
\langle \eta(t) \rangle =0,\qquad \langle \eta(t)\eta(t')\rangle=2\zeta T \delta(t-t').
\end{equation}

To numerically evolve the stochastic dynamics, we employ the Heun predictor-corrector scheme. For a discretized timestep $h$, the predictor step is first constructed as
\begin{equation}
\bar{S}_{n+1}=S_n+f(S_n)h+\eta_n,
\end{equation}
where
\begin{equation}
f(S)=-\frac{1}{\zeta}\frac{\partial G}{\partial S},
\end{equation}
and $\eta_n$ is sampled from a Gaussian distribution with variance proportional to $2\zeta T h$. The corrected value is then obtained through
\begin{equation}
S_{n+1}=S_n+\frac{h}{2}\left[f(S_n)+f(\bar{S}_{n+1})\right]+\eta_n.
\end{equation}

By numerically integrating the Langevin equation, we obtain stochastic trajectories for the entropy near equilibrium. The autocorrelation function is then computed from these trajectories, and the associated autocorrelation time is extracted from the exponential decay behaviour near criticality.

In addition to the Langevin analysis, we also compute the eigenvalues of the corresponding Fokker-Planck equation associated with the stochastic evolution on the free energy landscape. The probability distribution $P(S,t)$ for the order parameter evolves according to the partial differential equation given below \cite{Li,Li2}
\begin{equation}
\frac{\partial P(S,t)}{\partial t}=D \frac{\partial}{\partial S}\left\{e^{-\beta G(S)}\frac{\partial}{\partial S}\left[e^{\beta G(S)} P(S,t)\right]\right\},
\end{equation}
where $D=T/\zeta$ denotes the diffusion coefficient and $\beta=1/T$.

The stationary solution is given by
\begin{equation}
P_0 \propto e^{-\beta G(S)}.
\end{equation}
Introducing the spectral decomposition
\begin{equation}
P(S,t)=P_0^{1/2}\psi(S)e^{-\lambda t},
\end{equation}
the Fokker--Planck equation can be mapped to a Schr\"odinger-type eigenvalue equation of the form
\begin{equation}
-T\frac{d^2\psi}{dS^2}+V(S)\psi=\lambda \psi,
\end{equation}
where $V(S)$ is the effective potential determined by the generalized free energy landscape.

The numerical procedure employed for solving the eigenvalue equation follows the method outlined in Ref.~\cite{slowing}.

\section{Analytical derivation of the relaxation-time critical exponent}\label{appendix B}

In this appendix, we analytically derive the critical scaling behavior of the relaxation time near the second-order phase transition point. The derivation follows directly from the structure of the Gibbs free energy in the vicinity of the critical point.

Let the Gibbs free energy be expressed in terms of an order parameter $x$, which may correspond to the horizon radius for charged AdS black holes or the entropy in the Kerr-AdS case. Equilibrium configurations are determined from
\begin{equation}
\frac{\partial G(x,T)}{\partial x}=0.
\label{eq_eqm}
\end{equation}

At the critical point $(x_c,T_c)$, the free energy develops an inflection point structure. Consequently, the first few derivatives satisfy
\begin{equation}
\left.\frac{\partial G}{\partial x}\right|_c=0,\qquad\left.\frac{\partial^2 G}{\partial x^2}\right|_c=0,\qquad\left.\frac{\partial^3 G}{\partial x^3}\right|_c=0,
\label{critical_conditions}
\end{equation}
while the quartic derivative remains nonvanishing,
\begin{equation}
\left.\frac{\partial^4 G}{\partial x^4}\right|_c\neq 0.
\end{equation}

To analyze the near-critical behavior, we define the reduced temperature
\begin{equation}
\epsilon=\frac{T-T_c}{T_c},
\end{equation}
together with the deviation of the order parameter from its critical value,
\begin{equation}
\Delta x=x-x_c.
\end{equation}

Expanding the equilibrium condition (\ref{eq_eqm}) around the critical point yields
\begin{equation}
0=\left.\frac{\partial G}{\partial x}\right|_c+\left.\frac{\partial^2 G}{\partial x \partial T}\right|_c(T-T_c)+\frac{1}{3!}\left.\frac{\partial^4 G}{\partial x^4}\right|_c(\Delta x)^3+\cdots .
\label{expanded_eqm}
\end{equation}

Higher-order mixed terms involving both $\Delta x$ and $\epsilon$ are subleading near the critical point and are therefore neglected. Using the criticality conditions (\ref{critical_conditions}), the linear and quadratic terms in $\Delta x$ vanish identically. Retaining the leading contributions, one obtains 
\begin{equation}
A\,\epsilon+B\,(\Delta x)^3\simeq0,
\label{landau_balance}
\end{equation}
where
\begin{equation}
A=T_c\left.\frac{\partial^2 G}{\partial x \partial T}\right|_c,\qquad B=\frac{1}{3!}\left.\frac{\partial^4 G}{\partial x^4}\right|_c .
\end{equation}

Equation (\ref{landau_balance}) immediately implies
\begin{equation}
\Delta x\sim|\epsilon|^{1/3}.
\label{dx_scaling}
\end{equation}

The relaxation dynamics is governed by the curvature of the Gibbs free energy at equilibrium,
\begin{equation}
G''(x_e)=\left.\frac{\partial^2 G}{\partial x^2}\right|_{x=x_e}.
\end{equation}

Expanding this quantity near criticality gives
\begin{equation}
G''(x_e)=\left.\frac{\partial^2 G}{\partial x^2}\right|_c+\left.\frac{\partial^3 G}{\partial x^3}\right|_c\Delta x+\frac{1}{2}\left.\frac{\partial^4 G}{\partial x^4}\right|_c(\Delta x)^2+\cdots .
\end{equation}

Again, owing to the criticality conditions (\ref{critical_conditions}), the first two terms vanish identically, leaving
\begin{equation}
G''(x_e)\sim(\Delta x)^2.
\end{equation}

Using (\ref{dx_scaling}), one therefore obtains
\begin{equation}
G''(x_e)\sim|\epsilon|^{2/3}.
\label{gpp_scaling}
\end{equation}

The relaxation time is related to the inverse curvature of the free energy,
\begin{equation}
\tau\propto\frac{1}{G''(x_e)}.
\label{tau_inverse_curvature}
\end{equation}

Consequently,
\begin{equation}
\tau\sim|\epsilon|^{-2/3}.
\label{tau_final_scaling}
\end{equation}

Thus, the relaxation time diverges near the critical point with the universal exponent
\begin{equation}
\Delta=\frac{2}{3}.
\end{equation}

This result follows solely from the generic structure of the Gibbs free energy near a critical inflection point and is therefore independent of the microscopic details of the black hole background.

\end{document}